\documentclass[a4paper,12pt]{article}

\usepackage{amsmath,amssymb,array,amsfonts}
\usepackage{color}
\usepackage{geometry}
\usepackage[bf, small, hang]{caption}
\allowdisplaybreaks

\newcommand{\be}{\begin{equation}}
\newcommand{\ee}{\end{equation}}
\newcommand{\bea}{\begin{eqnarray}}
\newcommand{\eea}{\end{eqnarray}}

\makeatletter
    
    \@addtoreset{equation}{section}
\makeatother

\usepackage
[colorlinks=true,
 linkcolor=black,
 urlcolor=magenta,
 citecolor=cyan,
 hypertex]
 {hyperref}

\begin{document}
\setlength{\baselineskip}{18pt}
\begin{titlepage}

\begin{flushright}
OCU-PHYS 431 \\ 
KOBE-TH-15-09%
\end{flushright}
\vspace{1.0cm}
\begin{center}
{\Large\bf Predictions of the Higgs Mass and the Weak Mixing Angle \\
\vspace*{3mm}
in the 6D Gauge-Higgs Unification} 
\end{center}
\vspace{25mm}

\begin{center}
{\large
Kouhei Hasegawa$^{\,1}$, 
Chong-Sa Lim$^{\,2}$,
and Nobuhito Maru$^{\,3}$
}
\end{center}
\vspace{1cm}
\centerline{{\it
$^{1}$Department of Physics, 
Kobe University, Kobe 657-8501, Japan}}

\centerline{{\it 
$^{2}$The Department of Mathematics, Tokyo Woman's Christian University, Tokyo 167-8585, Japan}}  

\centerline{{\it
$^{3}$Department of Mathematics and Physics,
Osaka City University, Osaka 558-8585, Japan}}
%
%
\vspace{2cm}
\centerline{\large\bf Abstract}
\vspace{0.5cm}

In the gauge-Higgs unification with multiple extra spaces, 
the Higgs self-coupling is on the order of $g^2$ and the Higgs boson is predicted to be light, being consistent with the LHC results. 
When the gauge group is simple, the weak mixing angle is also predictable. 
We address a question on whether there exists a model of gauge-Higgs unification in 6-dimensional space-time, 
 which successfully predicts the mass ratios of the Higgs boson and weak gauge bosons. 
First, using a useful formula, we give a general argument on the condition for obtaining
a realistic prediction of the weak mixing angle $\sin^{2}\theta_{W} = 1/4$, 
and find that triplet and sextet representations of the minimal SU(3) gauge group lead to the realistic prediction. Concerning the Higgs mass, 
we notice that, in the models with one Higgs doublet, the predicted Higgs mass is always the same: $M_H = 2 M_W$. 
However, by extending our discussion to the models with two Higgs doublets, the situation changes: 
we obtain an interesting prediction $M_{H} \leq 2M_{W}$ at the leading order of the perturbation. 
Thus, it is possible to recover the observed Higgs mass, 125 GeV, for a suitable choice of the parameter. 
The situation is in clear contrast to the case of the minimal supersymmetric standard model, 
where $M_{H} \leq M_{Z}$ at the classical level and the predicted Higgs mass cannot recover the observed value.

\end{titlepage}




\newpage

\section{Introduction} 

The recent LHC data has revealed that the Higgs boson is ``light" with the mass of ${\cal O}(M_{W})$ \cite{ATLAS, CMS}. 
This implies that the Higgs self-coupling $\lambda$ is of ${\cal O}(g^{2}) \ (g: \ {\rm gauge \ coupling})$ and 
therefore governed by the gauge principle. 
Among various scenarios of physics beyond the standard model (BSM), 
the minimal supersymmetric standard model (MSSM) and gauge-Higgs unification (GHU) (formulated on multi-dimensional extra space) 
have such a desirable property and predict a light Higgs boson with  definite mass ratios of the Higgs boson to weak gauge bosons; 
i.e. $M_{H} \leq M_{Z} \cos 2\beta$ ($\beta$: an angle to denote the relative weight of the vacuum expectation values of the two Higgs doublets) for MSSM 
and $M_{H} = 2M_{W}$ for the 6-dimensional (6D) SU(3) GHU model with one Higgs doublet at the classical level \cite{SSSW, LMM}. 

Since these definite mass ratios are inevitable consequences of symmetries, i.e., SUSY and 
(higher-dimensional) local gauge symmetry, respectively, 
we naturally expect that even under quantum correction, deviations from the relations mentioned above are UV-finite and definitely predictable. 
In fact, in MSSM, it is well known that, under the quantum correction by the SUSY multiplet of the top quark 
$(t, \ \tilde{t})$, $M_{H}$ deviates from $M_{Z} \cos 2\beta $ due to the SUSY breaking, 
$m_{\tilde{t}}^{2} \gg m_{t}^{2}$, and the deviation is UV-finite and calculable.

In this paper, we focus on another scenario of BSM, i.e., GHU, 
where the Higgs boson originates from the extra space component of a higher dimensional gauge field \cite{1979Manton, 1983Hosotani} 
and therefore
the quantum correction to the Higgs mass-squared is finite due to the higher-dimensional local gauge symmetry, 
thus providing an alternative solution of the well-known gauge hierarchy problem \cite{1998HIL}. 
There have also been studies on the finiteness of the Higgs boson mass in the context of the gauge-Higgs unification in various models \cite{ABQ, finiteness}.

Interestingly, similarly to the case of MSSM, 
even under the quantum corrections, the Higgs mass itself and the deviation from the relation mentioned above, $M_{H} = 2M_{W}$, 
have been demonstrated to be both UV-finite in the 6D SU(3) GHU model with one Higgs doublet \cite{LMM}. 
In this case, although the local gauge symmetry still exists even after the compactification of the extra-space, 
the compactification to the non-simply connected extra space makes the Aharanov--Bohm (AB) phases along two different cycles of the torus 
physically meaningful and such nonlocal effects contribute to the finite deviation from the tree-level relations. 

Such predictability of the Higgs mass is desirable features of MSSM and GHU. 
However, the differences of the tree-level predictions of the Higgs mass from the observed value are rather large 
in both scenarios for the differences to be explained by quantum corrections:
\be 
\label{-1.1}
125 - M_{Z} \simeq 2M_{W} - 125 \simeq 35 {\rm GeV},  
\ee
where $M_{Z}$ is the maximum value of the tree level prediction of MSSM, while $2M_{W}$ is the prediction of the 6D SU(3) GHU 
model with one Higgs doublet. 
Thus, to realize the observed Higgs mass, i.e., $M_{H} = 125$ GeV, a considerably large SUSY breaking $M_{SUSY}$ or a considerably 
small bulk mass of matter field \cite{LMM} is required, respectively.

Hence, in this article, we address a question on whether there ever exist GHU models that provides more realistic tree level 
predictions of the Higgs mass. 
Namely, we investigate in the scheme of 6D GHU whether the tree level prediction of the Higgs mass becomes closer to or
coincides with the observed value of $125$ GeV, 
by suitable choices of the gauge group and the  compactification, especially the manner of orbifolding, 
which determines how many Higgs doublets of SU(2)$_L$ remain in the low energy effective theory as the KK zero modes. 

In the models of GHU, the gauge group of the standard model (SM) is forced to be enlarged, 
since the Higgs boson inevitably belongs to an adjoint representation (``repr.'' for short) of the gauge group in GHU, 
while in the SM, the Higgs boson belongs to the fundamental repr. of SU(2)$_L$. 
Thus, the minimal unified electro-weak model incorporating the SM is the SU(3) GHU model \cite{KLY, SSS}. 
In such models with the simple gauge group, the weak mixing angle, i.e.,
the mass ratio of weak gauge bosons, can also be predicted, in addition to the mass ratio of the Higgs to weak gauge bosons. 

Unfortunately, the predicted weak mixing angle in the minimal SU(3) model is far from the observed value: $\sin^{2}\theta_{W} = \frac{3}{4}$. 
Interestingly, however, it has been pointed out that a slightly larger gauge group G$_2$ leads to a successful prediction of the weak mixing angle: 
$\sin^{2}\theta_{W} = \frac{1}{4}$ \cite{1979Manton, CGM}. 

Thus, basically, we are in a position to predict mutual relations among all massive bosonic particles in the SM. 
The purpose of this article is to exploit the possibilities to realize realistic predictions on the Higgs mass 
and the weak mixing angle in the framework of the 6D GHU model with one or two Higgs doublets.

\section{Weak Mixing Angle and Representations under SU(3)} 

We first discuss the prediction of the weak mixing angle. 
We can demonstrate that the predicted weak mixing angle can be easily calculated without explicit calculations of the weak gauge 
boson masses $M_{W, Z}$, once we know the gauge group. 
More precisely, we will argue that, by knowing which repr. of the minimal group SU(3) the Higgs doublet belongs to, 
the weak mixing angle is immediately fixed. 
One reasonable assumption here is that the gauge group of GHU model includes SU(3) as its subgroup and the electro-weak gauge symmetry of the SM, 
SU(2)$_L \times$ U(1)$_Y$, is embedded into the simple group SU(3).  

A key formula in this argument is 
\be 
\label{0.1}
\sin^{2}\theta_{W} = \frac{{\rm Tr} \ I_{3}^{2}}{{\rm Tr} \ Q^{2}},  
\ee 
where ${\rm Tr} \ I_{3}^{2}$ and ${\rm Tr} \ Q^{2}$ are the summations of the squared eigenvalues of the operators $I_{3}$ and $Q$ 
(the charge operator in the unit of $e$) for an arbitrary repr. of SU(3). 
The proof of this useful relation (\ref{0.1}) is as follows. The essentially important relation is the orthogonality of the 
generators associated with photon and $Z$ boson, ${\rm Tr} \{ Q(I_{3} - \sin^{2}\theta_{W}Q)\} = {\rm Tr}(QI_{3}) - \sin^{2}\theta_{W}
{\rm Tr}Q^{2} = 0$, which holds generically for simple groups, since the gauge coupling is unique, and photon and $Z$ boson are two
orthogonal states. We also note that ${\rm Tr}(QI_{3}) = {\rm Tr}\{ (I_{3} + \frac{Y}{2})I_{3}\} = {\rm Tr}I_{3}^{2}$, where $Y$ denotes the
generator of the weak hypercharge and the orthogonality ${\rm Tr}(I_{3}Y) = 0$ has been used. We thus obtain 
${\rm Tr}I_{3}^{2} - \sin^{2}\theta_{W}{\rm Tr}Q^{2} = 0$, leading to $\sin^{2}\theta_{W} = \frac{{\rm Tr} \ I_{3}^{2}}{{\rm Tr} \ Q^{2}}$.

As the repr. of SU(3), we choose the simplest triplet. 
Since the triplet is decomposed under the subgroup SU(2)$_L$ as $3 \to  2 + 1$, 
the upper two components of the triplet can be regarded as the SU(2)$_L$ doublet, and therefore the electric charges of 
these upper two components differ by one unit. 
We also note that ${\rm Tr} \ Q = 0$, since the charge operator should be one of the generators of SU(3). 
Then, the charge assignment for the components of the triplet can be written generally in the form of 
\be 
\label{0.2} 
\begin{pmatrix} 
q \\ 
q-1 \\ 
1-2q
\end{pmatrix}, 
\ee 
with a parameter $q$. 
Then, using Eq.\,(\ref{0.1}), the weak mixing angle is written as 
\be 
\label{0.3} 
\sin^{2}\theta_{W}  = \frac{(\frac{1}{2})^{2} + (-\frac{1}{2})^{2} + 0}{q^{2}+(q-1)^{2}+(1-2q)^{2}} = 
\frac{1}{4(3q^{2}-3q+1)}.
\ee 

For instance, in the minimal SU(3) GHU model, the Higgs doublet inevitably belongs to the octet of SU(3). 
Since the octet is constructed by the product of the triplet and anti-triplet repr.s and the triplet is decomposed under the
subgroup SU(2)$_L$ as $2 + 1$, 
the neutral component of the Higgs doublet [\,SU(2) doublet\,] comes from the product of the second component [\,SU(2) doublet\,] 
and the complex conjugate of the third component [\,SU(2) singlet\,] of the triplet. 
[\,We may choose the second component, not the first one, without any loss of generality invoking the SU(2) symmetry\,]. 
Thus, the condition that the electric charge of the neutral Higgs boson vanishes is written as $q-1+ [-(1-2q)] = 3q-2 = 0$, 
leading to $q = \frac{2}{3}$. Thus, we get $\sin^{2}\theta_{W} = \frac{3}{4}$ from Eq.\,(\ref{0.3}) \cite{KLY, SSS}. 

When the adopted gauge group $G$ is larger than SU(3), its adjoint repr. generally contains various repr.s of the subgroup SU(3). 
For instance, in the case of $G = G_{2}$, the adjoint 14 repr. is decomposed under the subgroup SU(3) as $14 \to 8 + 3 + \bar{3}$. 
Thus, the Higgs doublet no longer has to belong to the octet, but may belong to other repr.s of SU(3). 
Namely, there appears a possibility of obtaining a realistic prediction of the weak mixing angle, $\sin^{2}\theta_{W} = \frac{1}{4}$. 
Note that (\ref{0.3}) implies that $\sin^{2}\theta_{W} = \frac{1}{4}$ is obtained if and only if $q = 1$ or 0. 

We find that the first possibility $q = 1$ to obtain $\sin^{2}\theta_{W} = \frac{1}{4}$ is realized if the Higgs doublet belongs 
to the triplet of SU(3). 
In fact, in this case, the second component of (\ref{0.2}) itself should be the neutral component, and $q-1 = 0 \ \to \ q = 1$. 
This is why the gauge group $G_{2}$ leads to $\sin^{2}\theta_{W} = \frac{1}{4}$ \cite{1979Manton, CGM}. 
To be more precise, in the case where the triplet component among $8 + 3 + \bar{3}$ develops the VEV, 
we obtain the desirable result, while if the octet develops the VEV, we again obtain $\sin^{2}\theta_{W} = \frac{3}{4}$, 
just as in the minimal SU(3) model. 

We point out that another new possibility $q = 0$ is realized if the Higgs doublet belongs to the 2nd-rank symmetric tensor 
repr., i.e., sextet repr. 6 of SU(3). 
Since the sextet is constructed by the symmetric product of two triplet repr.s, 
the neutral component of the Higgs doublet comes from the product of the second and third components of the triplet. 
Thus, the parameter $q$ is fixed as $q-1+ (1-2q) = -q = 0 \ \to \ q = 0$. 
In the next section, we discuss the Sp(6) GHU model, whose adjoint repr. is known to incorporate the sextet of SU(3), 
as the prototype model for realizing this new possibility. 

By discussing the repr.s 3, 6, and 8 of SU(3), we have exhausted all repr.s up to the 2nd-rank tensor. 
The argument is easily generalized. 
Suppose that the Higgs doublet belongs to a generic tensor repr. $R^{i_{1}, \cdots, i_{m}}_{\bar{i}_{1}, \cdots, \bar{i}_{\bar{m}}}$, 
where the indices $i$ and $\bar{i}$ denote the components of 3 and $\bar{3}$ of SU(3), respectively. 
The indices $i_{1}, \cdots, i_{m}$ and $\bar{i}_{1}, \cdots, \bar{i}_{\bar{m}}$ are supposed to be totally symmetrized, respectively. 
Then, it is easy to see that $q = 1$ is realized for $|m-\bar{m}| = 1$, 
whose simplest case is the triplet ($m = 1, \ \bar{m} = 0$) and $q = 0$ is realized for $|m-\bar{m}| = 2$, whose simplest case is the sextet ($m = 2, \ \bar{m} = 0$).  
      
Now, we know what repr.s of SU(3) lead to the realistic weak mixing angle, and this knowledge is useful for choosing the gauge group: 
we can focus on the gauge group whose adjoint repr. contains such desirable repr.s of SU(3). 
Once we have a model with the realistic weak mixing angle, the next step will be to investigate whether the model predicts a realistic Higgs mass at the same time. 
In the following sections, we address this question by taking several concrete models with one or two Higgs doublets in 6D space-time. 
Since the weak mixing angle crucially depends on the choice of the gauge group, 
it may be natural to expect that the mass ratio of the Higgs boson to the weak gauge boson also depends on the choice of the gauge group.  

What we discuss in the following two sections are models with gauge groups of rank 3. 
We do not discuss the $G_{2}$ model with the simpler group of rank 2, since it has already been shown that the model 
predicts $M_{H} = M_{Z}$ at the classical level, although the prediction of the weak mixing angle is 
realistic: $\sin^{2} \theta_{W} = 1/4$ \cite{CGM}.

\section{Sp(6) model} 

What we first discuss is the 6D Sp(6) GHU model with one Higgs doublet in its low energy effective theory. 
The reason for this choice is that the decomposition of the adjoint repr. of Sp(6) under its subgroup SU(3),   
\be 
\label{1.1} 
21 \to 8 + 6 + \bar{6} + 1, 
\ee 
contains 6 (or $\bar{6}$) repr. 
Thus, this is a prototype model to realize the realistic weak mixing angle $\sin^{2}\theta_{W} = 1/4$ by assigning the Higgs doublet 
to the sextet repr. of SU(3), 
the new possibility proposed in the previous section. 

Knowing that the prediction of the weak mixing angle is successful, 
the main purpose in this section is to study the ratio of the Higgs boson mass $M_{H}$ to the weak scale $M_{W}$, 
in addition to the concrete confirmation of the weak mixing angle. 
The ratio $M_{H}/M_{W}$ has been known to be 2 at the classical level in the 6D SU(3) GHU model with one Higgs doublet \cite{SSSW, LMM}. 
If the prediction of this ratio changes depending on the gauge group, as we have seen in the case of the weak mixing angle, 
there may be a chance of realizing a more realistic mass ratio $M_{H}/M_{W}$ in this Sp(6) model.

\subsection{Gauge kinetic term}  

The 21 generators of Sp(6) are given for the fundamental repr. 6 [\,$3 + \bar{3}$ under SU(3)\,] as follows: 
\bea
&&T^{a} = \frac{1}{2\sqrt{2}} 
\begin{pmatrix} 
\lambda^{a} & 0 \\ 
0 & - (\lambda^{a})^{\ast} 
\end{pmatrix} \ \ ({\rm for} \ a = 1 - 8), 
\label{2.1a} \\ 
&&T^{9} = \frac{1}{2\sqrt{2}} 
\begin{pmatrix} 
I & 0 \\ 
0 & -I 
\end{pmatrix},  
\label{2.1b} \\ 
&&T^{a} = \frac{1}{2\sqrt{2}} 
\begin{pmatrix} 
0 & M_{j} \\ 
M_{j} & 0 
\end{pmatrix} \ \ ({\rm for} \ a = 9+j, \ j =1 - 6),  
\label{2.1c} \\ 
&&T^{a} = \frac{1}{2\sqrt{2}} 
\begin{pmatrix} 
0 & -iM_{j} \\ 
iM_{j} & 0 
\end{pmatrix} \ \ ({\rm for} \ a = 15+j, \ j =1 - 6),  
\label{2.1d} 
\eea 
where $\lambda^{a}$ are Gell--Mann matrices and the six symmetric matrices $M_{j}$ are 
\bea 
&& M_{1} =  
\begin{pmatrix} 
\sqrt{2} & 0 & 0 \\ 
0 & 0 & 0 \\ 
0 & 0 & 0 
\end{pmatrix}, \ 
M_{2} =  
\begin{pmatrix} 
0 & 0 & 0 \\ 
0 & \sqrt{2} & 0 \\ 
0 & 0 & 0 
\end{pmatrix}, \ 
M_{3} =  
\begin{pmatrix} 
0 & 0 & 0 \\ 
0 & 0 & 0 \\ 
0 & 0 & \sqrt{2} 
\end{pmatrix}, 
\nonumber \\ 
&& 
M_{4} =  
\begin{pmatrix} 
0 & 1 & 0 \\ 
1 & 0 & 0 \\ 
0 & 0 & 0 
\end{pmatrix}, \ 
M_{5} =  
\begin{pmatrix} 
0 & 0 & 0 \\ 
0 & 0 & 1 \\ 
0 & 1 & 0 
\end{pmatrix}, \ 
M_{6} =  
\begin{pmatrix} 
0 & 0 & 1 \\ 
0 & 0 & 0 \\ 
1 & 0 & 0 
\end{pmatrix}. 
\label{2.2} 
\eea 
These generators satisfy an ortho-normal condition: 
\be 
\label{2.3} 
{\rm Tr} (T^{a}T^{b}) = \frac{1}{2} \delta_{ab}.  
\ee

We introduce 21 gauge fields $A^{a}_{M}$: 
\be 
\label{2.4} 
A_{M} \equiv \sum_{a = 1}^{21} A^{a}_{M} T^{a} = (A_{\mu}, \ A_{z}, \ A_{\bar{z}}),   
\ee 
where 
\be 
\label{2.5}
A_{z} \equiv \frac{A_{5} + i A_{6}}{\sqrt{2}}, \ \ A_{\bar{z}} \equiv (A_{z})^{\dagger} =  \frac{A_{5} - i A_{6}}{\sqrt{2}} \ \ 
\left(z \equiv \frac{x^{5} - i x^{6}}{\sqrt{2}} \right). 
\ee
By using the field strength tensor, 
\be 
\label{2.6} 
F_{MN} \equiv \partial_{M}A_{N} - \partial_{N}A_{M} - ig [A_{M}, A_{N}],   
\ee
the gauge kinetic term is constructed as usual: 
\be 
\label{2.7} 
-\frac{1}{2} {\rm Tr} (F^{MN}F_{MN}) = -\frac{1}{2} {\rm Tr} (F^{\mu \nu}F_{\mu \nu}) + 2 {\rm Tr} (F^{\mu} \ _{z}F_{\mu \bar{z}}) 
+ {\rm Tr} \{(F_{z \bar{z}})^{2}\},   
\ee 
where 
\bea 
&&F_{\mu z} = \partial_{\mu}A_{z} - \partial_{z}A_{\mu} - ig [A_{\mu}, A_{z}], \ \ 
F_{\mu \bar{z}} = (F_{\mu z})^{\dagger} = \partial_{\mu}A_{\bar{z}} - \partial_{\bar{z}}A_{\mu} - ig [A_{\mu}, A_{\bar{z}}],  
\label{2.8a} \\ 
&&F_{z \bar{z}} = \partial_{z}A_{\bar{z}} - \partial_{\bar{z}}A_{z} - ig [A_{z}, A_{\bar{z}}], 
\label{2.8b} 
\eea  
with 
\be 
\label{2.9}
\partial_{z} \equiv \frac{\partial_{5} + i \partial_{6}}{\sqrt{2}}, \ \ \partial_{\bar{z}} \equiv \frac{\partial_{5} - i \partial_{6}}{\sqrt{2}}. 
\ee

\subsection{Orbifolding and KK zero modes} 

In order to have one Higgs doublet as a KK zero-mode, we adopt an orbifold $T^{2}/Z_{6}$ as our extra space, 
imposing the invariance of the theory under $Z_{6}$ transformation: 
\be 
\label{2.10}
z \ \to \ \omega z \ \ \ (\omega^{6} = 1). 
\ee
The ``$Z_{6}$-parity" assignment for the fundamental 6 repr. is given by the matrix 
\be 
\label{2.11} 
P = {\rm diag} (\omega, \omega, \omega^{4}, \bar{\omega}, \bar{\omega}, \bar{\omega}^{4}). 
\ee 
Then, the corresponding $Z_{6}$-parities for 4D gauge and scalar fields are fixed as
\bea 
&&A_{\mu}(x^{\mu}, \omega z) =  P A_{\mu}(x^{\mu}, z) P^{\dagger}, 
\label{2.12a} \\ 
&&A_{z}(x^{\mu}, \omega z) =  \omega P A_{z}(x^{\mu}, z) P^{\dagger}, 
\label{2.12b} \\ 
&&A_{\bar{z}}(x^{\mu}, \omega z) =  \bar{\omega} P A_{\bar{z}}(x^{\mu}, z) P^{\dagger}. 
\label{2.12c} 
\eea
 
We thus realize that the KK zero-modes of 4D gauge bosons are those of SU(2)$_L \times$ U(1)$_Y$, together with an 
additional U(1) gauge boson, and the KK zero-modes of 4D scalars just correspond to our Higgs doublet, $H = (\phi^{+}, \phi^{0})^{t}$: 
\bea 
&&A_{\mu} =  
\begin{pmatrix} 
a_{\mu} & 0 \\ 
0 & -a_{\mu}^{\ast} 
\end{pmatrix} + A^{9}_{\mu}T^{9}, 
\label{2.13a} \\ 
&&A_{z} =  
\begin{pmatrix} 
0 & a_{z} \\ 
0 & 0 
\end{pmatrix}, 
\label{2.13b} 
\eea 
where the $3 \times 3$ matrices $a_{\mu}$ and $a_{z}$ are given as 
\bea 
&&a_{\mu} =  
\begin{pmatrix} 
\frac{\sqrt{6}}{6}Z_{\mu} & \frac{1}{2} W^{+}_{\mu} & 0 \\ 
\frac{1}{2} W^{-}_{\mu} & -\frac{\sqrt{2}}{4}\gamma_{\mu} - \frac{\sqrt{6}}{12}Z_{\mu} & 0 \\ 
0 & 0 & \frac{\sqrt{2}}{4}\gamma_{\mu} - \frac{\sqrt{6}}{12}Z_{\mu} 
\end{pmatrix}  
\label{2.14a} \\ 
&&a_{z} = \frac{1}{2} 
\begin{pmatrix} 
0 & 0 &  \phi^{+} \\ 
0 & 0 &  \phi^{0} \\ 
\phi^{+} & \phi^{0} & 0 
\end{pmatrix}. 
\label{2.14b} 
\eea 
$\gamma_{\mu}$ and $A^{9}_{\mu}$ stand for the photon and the extra U(1) gauge boson, respectively.
Note that the photon field appears in (\ref{2.14a}) so that the coupled charge operator is written as 
\be 
\label{2.15}
Q = {\rm diag} (0, \ -1, \ 1) = \frac{1}{2}\lambda^{3} + \frac{\sqrt{3}}{2}(-\lambda^{8}), 
\ee 
whose form is fixed by the condition that $\phi^{0}$ in (\ref{2.14b}) is electrically neutral. 
(\ref{2.15}) in turn implies that $\sin \theta_{W} = \frac{1}{2}$ and $\cos \theta_{W} = \frac{\sqrt{3}}{2}$, and therefore 
\be 
\label{2.17}
\sin^{2} \theta_{W} = \frac{1}{4}, 
\ee
as we expected.

\subsection{Mass ratios of weak gauge bosons and Higgs boson} 

In this subsection, we calculate the masses of the weak gauge bosons $W^{\pm}_{\mu}$ and $Z_{\mu}$, and 
the Higgs boson $h \ [\,\phi^{0} = \frac{v + h +iG^{0}}{\sqrt{2}}$ with $v$ being the vacuum expectation value (VEV) of the Higgs field\,]. 
For that purpose, we need $\kappa, \ \kappa'$, and $\lambda$ defined as the coefficients of the relevant part of the lagrangian \cite{LMM},  
\be 
\label{2.19}
\kappa |\phi^{(0)}|^{2}W^{+ \mu}W^{-}_{\mu} + \kappa' |\phi^{(0)}|^{2}Z^{\mu}Z_{\mu} - \lambda |\phi^{(0)}|^{4}. 
\ee
Note that, in GHU, the quadratic term of the Higgs field does not exist at the tree level and is induced at the quantum level 
with a UV-finite coefficient \cite{LMM}. 
Once the VEV $v$ is generated by the radiatively induced negative mass-squared term, 
the masses of $W^{\pm}_{\mu}, \ Z_{\mu}$, and $h$ can be written in terms of these coefficients as  
\be 
\label{2.20}
M_{W}^{2} = \frac{\kappa}{2}v^{2}, \ M_{Z}^{2} = \kappa' v^{2}, \ M_{H}^{2} = 2\lambda v^{2}.  
\ee 

The coefficients $\kappa, \ \kappa'$, and $\lambda$ can be read off from the commutator squared 
in ${\rm Tr} (F^{\mu} \ _{z}F_{\mu \bar{z}})$ and ${\rm Tr} \{(F_{z \bar{z}})^{2}\}$: 
\bea 
&& 2 {\rm Tr} (F^{\mu} \ _{z}F_{\mu \bar{z}}) \ \ \to \ \ -2g^{2}{\rm Tr} \{[A^{\mu}, A_{z}][A_{\mu}, A_{\bar{z}}]\} 
= g^{2}|\phi^{0}|^{2} \left( \frac{1}{4}W^{+\mu}W^{-}_{\mu} + \frac{1}{6}Z^{\mu}Z_{\mu} \right),  
\label{2.22a} \\ 
&& {\rm Tr} \{(F_{z \bar{z}})^{2}\} \ \ \to \ \ -g^{2}{\rm Tr} \{ [A_{z}, A_{\bar{z}}]^{2} \} = - \frac{g^{2}}{4}|\phi^{0}|^{4}. 
\label{2.22b} 
\eea 
We find 
\be
\label{2.23}
\kappa = \frac{g^{2}}{4}, \ \kappa' = \frac{g^{2}}{6}, \ \lambda = \frac{g^{2}}{4}, 
\ee 
which in turn mean, from (\ref{2.20}),   
\be 
\label{2.24}
M_{W}^{2} = \frac{g^{2}}{8}v^{2}, \ M_{Z}^{2} = \frac{g^{2}}{6} v^{2}, \ M_{H}^{2} = \frac{g^{2}}{2}v^{2}.  
\ee 
We thus conclude that
\be 
\label{2.25}
M_{W} = \frac{\sqrt{3}}{2}M_{Z}, \ \ 
M_{H} = 2 M_{W}. 
\ee 
The former relation is consistent with $\sin^{2} \theta_{W} = 1/4 \ \left( \rho = \frac{M_{W}^{2}}{M_{Z}^{2}\cos^{2} \theta_{W}} = 1 \right)$. 
The latter relation, however, is exactly the same as that predicted in the SU(3) model with one Higgs doublet 
($Z_{3}$ orbifolding) \cite{SSSW, LMM}, 
and unfortunately, we cannot obtain a closer Higgs mass to the observed value by adopting Sp(6).

\section{SU(4) model} 

We have already mentioned that the exceptional group $G_{2}$ leads to the realistic weak mixing angle $\sin^{2} \theta_{W} = 1/4$, 
if the Higgs doublet belongs to the triplet component of the subgroup SU(3). 
There is another familiar gauge group, whose adjoint repr. contains the triplet, i.e., SU(4): 
the adjoint repr. 15 is decomposed under SU(3) as $15 \to 8 + 3 + \bar{3} + 1$. 
In this section, we address a question whether or not this another possibility of the gauge group predicts a desirable Higgs mass. 
We just follow the argument made in the previous section and will skip the detail. 

The orbifold we adopt is $T^{2}/Z_{6}$. The ``$Z_{6}$-parity" assignment for the fundamental repr. is given by a $4 \times 4$ matrix 
\be 
\label{2'.1}  
P = {\rm diag} (1, 1, \omega^{3}, \omega) \ \ (\omega^{6} = 1). 
\ee 
Accordingly, the KK zero-modes for 4D gauge and scalar fields are written as   
\bea 
&&A_{\mu} =  
\begin{pmatrix} 
\frac{1}{2}\gamma_{\mu} - \frac{\sqrt{3}}{6}Z_{\mu} & \frac{1}{\sqrt{2}} W^{+}_{\mu} & 0 & 0 \\ 
\frac{1}{\sqrt{2}} W^{-}_{\mu} & \frac{\sqrt{3}}{3}Z_{\mu} & 0 & 0 \\ 
0 & 0 & - \frac{1}{2}\gamma_{\mu} - \frac{\sqrt{3}}{6}Z_{\mu} & 0 \\ 
0 & 0 & 0 & 0 
\end{pmatrix} + {\rm the \ extra \ U(1) \ gauge \ boson}, 
\label{2'.2a} \\ 
&&A_{z} =  \frac{1}{\sqrt{2}} 
\begin{pmatrix} 
0 & 0 & 0 & \phi^{+} \\ 
0 & 0 & 0 & \phi^{0} \\ 
0 & 0 & 0 & 0 \\ 
0 & 0 & 0 & 0  
\end{pmatrix}. 
\label{2'.2b} 
\eea 
(\ref{2'.2b}) means that the Higgs doublet behaves as a triplet repr. of the subgroup SU(3). 

Again, the coefficients $\kappa, \ \kappa'$, and $\lambda$ of (\ref{2.19}) can be read off from the commutator-squared in 
$\rm{Tr} (F^{\mu} \ _{z}F_{\mu \bar{z}})$ and $\rm{Tr} \{(F_{z \bar{z}})^{2}\}$: 
\bea 
&& 2 \rm{Tr} (F^{\mu} \ _{z}F_{\mu \bar{z}}) \ \ \to \ \  2g^{2}|\phi^{0}|^{2} \left( \frac{1}{4}W^{+\mu}W^{-}_{\mu} + \frac{1}{6}Z^{\mu}Z_{\mu} \right),  
\label{2'.3a} \\ 
&& \rm{Tr} \{(F_{z \bar{z}})^{2}\} \ \ \to \ \ - \frac{g^{2}}{2}|\phi^{0}|^{4}. 
\label{2'.3b} 
\eea 
Thus, we conclude that 
\be
\label{2'.4}
\kappa = \frac{g^{2}}{2}, \ \kappa' = \frac{g^{2}}{3}, \ \lambda = \frac{g^{2}}{2}, 
\ee 
which in turn mean that   
\be 
\label{2'.5}
M_{W}^{2} = \frac{g^{2}}{4}v^{2}, \ M_{Z}^{2} = \frac{g^{2}}{3} v^{2}, \ M_{H}^{2} = g^{2}v^{2}.  
\ee 
We realize that, although the weak mixing angle is realistic, $M_{W} = \frac{\sqrt{3}}{2}M_{Z} \ (\sin^{2} \theta_{W} = 1/4)$, 
the predicted Higgs mass $M_{H} = 2M_{W}$ is again the same as in the cases of the SU(3) and Sp(6) models.

\section{SU(3) Model with Two Higgs Doublets} 

So far, we have studied 6D GHU models with only one Higgs doublet in its low energy effective theory and have seen 
that all the models predict $M_{H} = 2M_{W}$, 
which is rather far from the observed value of $M_{H}$ at LHC experiments for the quantum correction to recover the difference. 
Now, we consider the possibility of realizing a prediction of $M_{H}$, 
which is closer to or even coincides with the observed value, in the framework of the 6D GHU model with two Higgs doublets. 
It is interesting to note that the MSSM has some similarity to such a GHU model, 
having two Higgs doublets and the Higgs self-coupling being governed by the gauge principle; $\lambda \sim g^{2}$. 
In MSSM, however, the tree level prediction is $M_{H} \leq M_{Z} \cos 2\beta$ and there is no chance for the tree level 
prediction to coincide with the observed value.  

One remark here is that, in our model, the quadratic terms of the Higgs doublets do not exist at the tree level, 
while the quartic self-coupling is provided by $g^{2}[A_{5}, A_{6}]^{2}$ of the gauge kinetic term at the tree level, 
as we have seen in the previous sections. 
Thus, we are going to calculate the 1-loop induced quadratic terms. 
Our attitude here is to consider only the leading contribution of the perturbative expansion to each term of the Higgs potential. 

\subsection{The model} 
The model of interest is a 6D SU(3) GHU model with an orbifold $T^{2}/Z_{2}$ as its extra space. 
The $Z_{2}$ orbifolding is needed to obtain a chiral theory and the necessary breaking SU(3) $\to$ SU(2)$_L \times$ U(1)$_Y$, 
but we still have two Higgs doublets coming from the two extra space components $A_{5}$ and $A_{6}$ 
[\,For instance, the $Z_{3}$ orbifolding leaves only one Higgs doublet at the KK zero-mode sector \cite{SSSW, LMM}\,]. 

The torus $T^{2}$ are described by extra space coordinates $(x^{5}, x^{6})$. 
For simplicity, lattice vectors along the two independent cycles of the torus $\vec{l}_{1,2}$ are assumed to satisfy  
\be 
\label{3.1} 
|\vec{l}_{1}| = |\vec{l}_{2}| = 2\pi R, \ \ \vec{l}_{1} \perp \vec{l}_{2}.  
\ee 
The $Z_2$-parity assignment for the triplet of SU(3) is given by  
\be 
\label{3.3}
P = {\rm diag} (1, 1, -1).
\ee 
Accordingly, the $Z_2$-parities for the gauge-Higgs sector are fixed as 
\bea 
A_{\mu}(-x^{5}, -x^{6}) &=& PA_{\mu}(x^{5}, x^{6})P^{-1}, \nonumber \\ 
A_{5, 6}(-x^{5}, -x^{6}) &=& - PA_{5, 6}(x^{5}, x^{6})P^{-1}. 
\label{3.4}
\eea
We thus realize that the KK zero-modes of 4D gauge bosons $A_{\mu}$ are just those of SU(2)$_L \times$ U(1)$_Y$, 
although the predicted weak mixing angle is unrealistic: $\sin^{2}\theta_{W} = 3/4$ 
[\,We will consider SU(4) model in the next section to evade this problem\,]. 
As the KK zero-modes of 4D scalars $A_{5, 6}$, we obtain two Higgs doublets $H_{1, 2}$:  
\be 
\label{3.5} 
A_{5, 6}^{(0, 0)} =  \frac{1}{\sqrt{2}} 
\begin{pmatrix} 
0 & 0 & \phi_{1,2}^{+} \\ 
0 & 0 & \phi_{1,2}^{0} \\ 
\phi_{1,2}^{-} & \phi_{1,2}^{0\ast} & 0 
\end{pmatrix},  
\ee 
with 
\be 
\label{3.5'}
H_{1,2} = 
\begin{pmatrix} 
\phi_{1,2}^{+} \\ 
\phi_{1,2}^{0} 
\end{pmatrix}.
\ee

\subsection{A general analysis of the Higgs potential and Higgs mass} 

Here, we study the Higgs mass in the two Higgs doublet model in a general framework of the effective Higgs potential, 
where the quadratic term of the Higgs fields is assumed to take a general form allowed by gauge invariance, 
while the quartic self-coupling term is given by its tree level contribution: 
\be 
\label{3.6}
{\rm Tr} \{(F_{5 6})^{2}\} \ \ \to \ \  - g^{2} {\rm Tr} \{ [A_{5}, A_{6}]^{2} \}. 
\ee 

Let us note that there should be no local operators responsible for the quadratic terms of the Higgs fields, 
since the gauge invariance in the bulk implies that the gauge-invariant and Lorentz-invariant local operators are written 
by the use of field strength, and even the operator with minimum mass dimension, $F_{MN}F^{MN}$, already has mass 
dimension 4 (from 4D   viewpoint) 
and therefore does not contain the quadratic terms. 
Thus, the only possible operators relevant to the quadratic terms are either global operators due to 
the Wilson-loops along two independent cycles of the torus,  
\be 
\label{3.7} 
P \{ e^{ig \oint A_{5, 6} \ dy_{1,2}} \}, 
\ee
or the ``tadpole" term, the linear term of $F_{56}$ corresponding to U(1)$_{Y}$ localized at the fixed points of the orbifold, 
which leads to a quadratic term ${\rm Tr} (Y [A_{5}, A_{6}])$ [\,$Y = {\rm diag} (\frac{1}{3}, \frac{1}{3}, - \frac{2}{3})$ is
the U(1)$_Y$ generator\,]. 
The tadpole term is consistent with the remaining gauge symmetry at the fixed points, SU(2)$_{L}\times$ U(1)$_{Y}$, although it
contradicts with the bulk gauge symmetry. 
This possible tadpole term, being a local operator, may be induced together with a UV-divergent coefficient. 

Thus, a general form of the effective potential with respect to the two Higgs doublets up to the quartic term is written as 
\be 
\label{3.8} 
V(H_{1}, H_{2}) = - \lambda \ {\rm Tr} \{ [A_{5}^{(0, 0)}, A_{6}^{(0, 0)}]^{2} \} + a \ {\rm Tr} [(A_{5}^{(0, 0)})^{2} +
(A_{6}^{(0, 0)})^{2}] + i b \ {\rm Tr}\{ Y [A_{5}^{(0, 0)}, A_{6}^{(0, 0)}]\}.    
\ee 
Among the quadratic terms, the term with the coefficient $a$ is expected to come from the Wilson-loops (\ref{3.7}), 
while the term with the coefficient $b$ is expected to be the contribution of the tadpole. 
The potential (\ref{3.8}) is 4-dimensional and is written in terms of the KK zero-modes $A_{5, 6}^{(0, 0)}$, 
since the Wilson-loops (\ref{3.7}) obtain contributions only from the KK zero-modes. 
We have assumed that the coefficients of $(A_{5}^{(0, 0)})^{2}$ and $(A_{6}^{(0, 0)})^{2}$ are the same, 
since we have assumed $|\vec{l}_{1}| = |\vec{l}_{2}|$ for the torus [\,see (\ref{3.1})\,]. 
The concrete form of (\ref{3.8}) in terms of two Higgs doublets $H_{1,2}$ is calculated to be 
\bea 
V(H_{1}, H_{2}) &=& \frac{\lambda}{2} \{ (H_{1}^{\dagger}H_{1})(H_{2}^{\dagger}H_{2}) 
+ (H_{1}^{\dagger}H_{2})(H_{2}^{\dagger}H_{1}) - (H_{2}^{\dagger}H_{1})^{2} - (H_{1}^{\dagger}H_{2})^{2} \} \nonumber \\ 
&&+ a (H_{1}^{\dagger}H_{1} + H_{2}^{\dagger}H_{2}) - \frac{i}{2}b (H_{1}^{\dagger}H_{2} - H_{2}^{\dagger}H_{1}).     
\label{3.8'} 
\eea 

At the tree level, 
\be 
\label{3.9} 
\lambda = g^{2}, \ \ a = b = 0. 
\ee
Thus, the leading contributions to the quadratic terms appear at the 1-loop level, and we will calculate the quantum corrections later. 

\subsubsection{The minimization} 

Here, supposing $a$ and $b$ are radiatively induced, let us perform the minimization of the potential and calculate the mass 
eigenvalues of 4D scalar particles. 

First, we find that  
\be 
\label{3.9'}
a > 0  
\ee 
is necessary, otherwise the potential becomes unstable in the ``flat-direction", where $H_{1} \propto H_2$ and the quartic 
term disappears: $[A_{5}^{(0, 0)}, A_{6}^{(0, 0)}] = 0$. 

For the purpose of minimization, we first focus on the neutral components $\phi_{1,2}^{0}$ of the doublets. Then the potential (\ref{3.8'}) reduces to  
\be 
\label{3.10} 
V (\phi_{1}^{0}, \phi_{2}^{0}) = 2 \lambda \ \{{\rm Im} (\phi_{1}^{0\ast} \phi_{2}^{0})\}^{2} + a \ (|\phi_{1}^{0}|^{2} + |\phi_{2}^{0}|^{2}) +  b \ {\rm Im} (\phi_{1}^{0\ast} \phi_{2}^{0}).    
\ee
In terms of the VEVs of the neutral components, 
\be 
\label{3.11} 
|\phi_{1, 2}^{0}| = \frac{v_{1, 2}}{\sqrt{2}}, \ \ \phi_{1}^{0\ast}\phi_{2}^{0} = \frac{v_{1}v_{2}}{2}e^{-i\theta} \ \ (v_{1,2} \geq 0),  
\ee 
the potential reads as 
\be 
\label{3.12}
V(v_{1}, v_{2}, \theta ) = \frac{\lambda}{2} (v_{1}v_{2} \sin \theta)^{2} + \frac{a}{2}(v_{1}^{2} + v_{2}^{2}) - \frac{b}{2} v_{1}v_{2} \sin \theta.  
\ee 

The minimization goes as follows. 
Let $x \equiv v_{1}v_{2} > 0, \ y \equiv v_{1} - v_{2}$. 
Then, we can complete the square: 
\be 
\label{3.14}
V = \frac{\lambda \sin^{2}\theta}{2} \Bigl(x - \frac{b \sin \theta - 2a}{2\lambda \sin^{2}\theta} \Bigr)^{2} + \frac{a}{2}y^{2} 
- \frac{(b -  \frac{2a}{\sin \theta})^{2}}{8\lambda}.  
\ee 
We need to impose 
\be 
\label{3.15}
b \sin \theta - 2a > 0 \ \ \to \ \ |b| > 2a,  
\ee 
since otherwise the minimum of the potential is at $x = y = 0$, i.e., at $v_{1, 2} = 0$, and there is no spontaneous symmetry breaking. 
Under the condition (\ref{3.15}), we easily realize that the minimum is at $b \sin \theta = |b|$. 
For instance, for $b > 0$, (\ref{3.15}) implies $\sin \theta > \frac{2a}{b} > 0$ and the vacuum energy $-(b -  \frac{2a}{\sin \theta})^{2}/8\lambda$ takes its minimum at $\sin \theta = 1$. 
We thus replace $\sin^{2}\theta \to 1, \ b \sin \theta \to |b|$ in (\ref{3.14}) to obtain 
\be 
\label{3.16}
V = \frac{\lambda}{2} \Bigl(x - \frac{|b| - 2a}{2\lambda} \Bigr)^{2} + \frac{a}{2}y^{2} 
- \frac{(|b| -  2a)^{2}}{8\lambda}. 
\ee 
Thus, we finally obtain the VEVs of the Higgs fields,  
\be 
\label{3.17} 
x = \frac{|b| - 2a}{2\lambda}, \ y = 0 \ \ \to \ \ v_{1} = v_{2} = \frac{v}{\sqrt{2}} = \sqrt{\frac{|b| - 2a}{2\lambda}} \ \ 
\left(\theta = \epsilon(b) \frac{\pi}{2} \right), 
\ee 
where $\epsilon (b)$ is the sign-function of $b$: $\epsilon (b) = \pm 1$ depending on the sign of $b$. In (\ref{3.17}), 
$v$ should be understood as the VEV corresponding to that in the SM, since the mass of the charged weak gauge boson is given in this model as 
\be 
\label{3.18}
M_{W}^{2} = \frac{g^{2}}{4}(v_{1}^{2} + v_{2}^{2}) = \frac{g^{2}}{4}v^{2}.  
\ee 

\subsubsection{Mass eigenvalues of 4D scalars} 

We will have five physically remaining scalar particles, just as in the MSSM. 
We now derive the mass eigenvalues of these physical states. 
It will be useful to perform a unitary transformation between two Higgs doublets, so that only one doublet develops the VEV $v$: 
\bea 
&& H \equiv \frac{1}{\sqrt{2}} [H_{1} + i\epsilon (b) H_{2}],  \nonumber \\ 
&& \tilde{H} \equiv \frac{1}{\sqrt{2}} [H_{1} - i\epsilon (b) H_{2}].  
\label{3.19}
\eea
From (\ref{3.17}), we realize that only $H$ develops a nonvanishing VEV: 
\be 
\label{3.20} 
\langle H \rangle = 
\begin{pmatrix} 
0 \\ 
\frac{v}{\sqrt{2}}
\end{pmatrix}, \ \ \langle \tilde{H} \rangle = 
\begin{pmatrix} 
0 \\ 
0
\end{pmatrix}, 
\ee
where we have assumed that the VEV of $H$ is real without loss of generality. 
Namely, $H$ is regarded as the doublet behaving as one of the SM.

Then, the NG bosons $G^{\pm}$ and $G^{0}$ and the physical Higgs scalar fields are denoted as 
\be 
\label{3.21}
H = 
\begin{pmatrix} 
G^{+} \\ 
\frac{v + h + iG^{0}}{\sqrt{2}}
\end{pmatrix}, \ \ 
\tilde{H} = 
\begin{pmatrix} 
h^{+} \\ 
\frac{\tilde{h} + iP}{\sqrt{2}}
\end{pmatrix}. 
\ee
Note that the NG bosons should belong to the doublet developing the VEV, i.e., $H$, and do not mix with physical 
scalar fields at the mass-squared matrices. 
That is why this base is convenient for analyzing the mass eigenvalues. 

Our task is to calculate the mass eigenvalues for the physical states, $h^{+}$ and $P$, and especially the ``CP-even" neutral 
Higgs $h$ and $\tilde{h}$. 
For that purpose, we rewrite the Higgs potential (\ref{3.8'}) by the use of $H$ and $\tilde{H}$: 
\bea 
V(H, \tilde{H}) &=& \frac{\lambda}{2} [(H^{\dagger}H)^{2} + (\tilde{H}^{\dagger}\tilde{H})^{2} 
- (H^{\dagger}H)(\tilde{H}^{\dagger}\tilde{H}) - (H^{\dagger}\tilde{H})(\tilde{H}^{\dagger}H)] \nonumber \\ 
&&+ a (H^{\dagger}H + \tilde{H}^{\dagger}\tilde{H}) - \frac{|b|}{2} (H^{\dagger}H - \tilde{H}^{\dagger}\tilde{H}).    
\label{3.22} 
\eea 
Substituting (\ref{3.21}) in (\ref{3.22}) and extracting the quadratic terms of the fields, we get  
\bea 
V(H, \tilde{H})_{{\rm quadratic}} &=& 0 \times |G^{+}|^{2} + (\frac{3}{2}a + \frac{1}{4}|b|) \ |h^{+}|^{2} \nonumber \\ 
&&+ \ 0 \times (G^{0})^{2} + \frac{1}{2}(2a) \ P^{2} + \frac{1}{2}(|b| - 2a) \ h^{2} + \frac{1}{2}(2a) \ \tilde{h}^{2},     
\label{3.23} 
\eea 
where the relation 
\be 
\label{3.24} 
\lambda v^{2} = |b| - 2a,  
\ee
obtained from (\ref{3.17}) has been used to show that $G^{\pm}$ and $G^{0}$ have vanishing masses. 
Namely, in the base of $H$ and $\tilde{H}$, the mass-squared matrix has been automatically diagonalized. 
In particular, there is no freedom of the angle $\alpha$ in the MSSM in our model owing to the adopted simplified torus compactification. 
Thus, as we expected, $G^{\pm}$ and $G^{0}$ are NG bosons and the masses of the remaining five physical states are given as 
\bea 
{\rm charged \ sector}&:& \ \ M_{h^{+}}^{2} = \frac{3}{2}a + \frac{1}{4}|b| = 2a + \frac{1}{4}\lambda v^{2} = 2a + M_{W}^{2}, \nonumber \\ 
\mbox{CP-odd sector}&:& \ \ M_{P}^{2} = 2a, \nonumber \\ 
\mbox{CP-even sector}&:& \ \ M_{h}^{2} = |b| - 2a = \lambda v^{2} = (2M_{W})^{2}, \ \ M_{\tilde{h}}^{2} = 2a,  
\label{3.25}
\eea 
where the relations (\ref{3.9}), (\ref{3.18}), and (\ref{3.24}) have been used.

If we identify the lighter CP-even neutral scalar with our Higgs particle, its mass is given depending on the relative 
magnitude of $a$ to $2M_{W}^{2}$ as 
\be 
\label{3.53} 
M_{H} = 
\begin{cases}  
2M_{W} & \text{(for $a > 2M_{W}^{2}$)} \\ 
\sqrt{2a} & \text{(for $a < 2M_{W}^{2}$)}. 
\end{cases}
\ee 

Interestingly, for $a < 2M_{W}^{2}$, the Higgs mass is predicted to be $M_{H} < 2M_{W}$ and even coincides with
the observed value $M_{H} = 125$ GeV for a choice of $a$. 
We should note also that, in this case, the mass of the CP-odd neutral scalar $P$ is degenerated with $M_{H}$, 
while the mass of the charged Higgs $\sqrt{2a + M_{W}^{2}}$ is larger than $M_{H}$ but on the order of the weak scale, 
which may be potentially dangerous when confronted with the LHC data. 
Fortunately, however, the present experimental lower bounds for the masses of the exotic scalar particles are rather loose: 
even for the charged scalar, the lower bound is still around 80 GeV or so \cite{data}. 
Note that, in this case, the lighter Higgs $\tilde{h}$ does not belong to the doublet developing the VEV, 
although it should have Yukawa couplings with fermions through its higher-dimensional gauge interaction.  

In the opposite case of $a > 2M_{W}^{2}$, the Higgs mass is just $2M_{W}$ and we recover 
the prediction of the one Higgs doublet models. 
Now, the Higgs belongs to the doublet developing the VEV $v$, just as in the SM. 
In this case, as long as the coefficient $a$ is sufficiently large, the other physical scalars become massive. 
In fact, in the limit of the compactification scale $M_{c} \equiv \frac{1}{R} \to \infty$,
we expect that the theory reduces to the SM, 
and such decoupling of four additional scalars, $h^{\pm}, \ P$, and $\tilde{h}$, and therefore the recovery of 
the prediction of the one doublet model are reasonable. 
We will see that $a$ is 
one order of magnitude smaller than
${\cal O} (\alpha M_{c}^{2}) \ [\,\alpha: {\rm the \ fine \ structure \ constant}$, see (\ref{4.20}) and (\ref{5.10})\,]. 
This situation mimics the case of MSSM, where in the limit $M_{SUSY} \to \infty$ the SM is expected to be recovered. 

To summarize, the 6D GHU model with two Higgs doublets has a desirable feature that it predicts 
\be 
\label{3.54}
M_{H} \leq 2M_{W}, 
\ee
and the predicted Higgs mass may even coincide with the observed value already at the leading order of the perturbative expansion. 
This is in clear contrast to the case of MSSM, where $M_{H} \leq M_{Z} \cos 2\beta \ (\leq M_{Z})$ at the leading order 
and cannot agree with the observed value.  

In order to confirm that the quadratic terms really take the form shown in (\ref{3.8'}), 
derived from an argument relying on the gauge symmetry and the symmetry of the torus, 
we now perform concrete calculations of the quantum corrections to the quadratic terms in two 
types of models: a model with a matter scalar and a model with a matter fermion.

\subsection{A model with matter scalar} 

We introduce SU(3) triplet complex scalar fields to the theory as the matter fields: 
\be 
\label{4.1} 
\Phi = 
\begin{pmatrix} 
\varphi_{1} \\ 
\varphi_{2} \\ 
\varphi_{3} 
\end{pmatrix}. 
\ee 
The $Z_{2}$-parity assignment for this matter fields is 
\be 
\label{4.2} 
\Phi (-x^{5}, -x^{6}) = P \Phi (x^{5}, x^{6}),  
\ee 
where $P$ is given in (\ref{3.3}). 
Thus, its KK mode expansion is
\be 
\label{4.3}
\Phi (x^{\mu}, x^{5}, x^{6}) = \sum_{n_{1}=-\infty}^{\infty} \sum_{n_{2}=-\infty}^{\infty} \frac{1}{2 \pi R} 
\begin{pmatrix} 
\cos ( \frac{n_{1}x^{5} + n_{2}x^{6}}{R} ) \varphi_{1}^{(n_{1}, n_{2})}(x^{\mu}) \\ 
\cos ( \frac{n_{1}x^{5} + n_{2}x^{6}}{R} ) \varphi_{2}^{(n_{1}, n_{2})}(x^{\mu}) \\ 
i \sin ( \frac{n_{1}x^{5} + n_{2}x^{6}}{R} ) \varphi_{3}^{(n_{1}, n_{2})}(x^{\mu}) 
\end{pmatrix},  
\ee 
where there are degeneracies among the 4D fields because of the $Z_{2}$ orbifolding:   
\bea 
&&\varphi_{1, 2}^{(-n_{1}, -n_{2})}(x^{\mu}) = \varphi_{1, 2}^{(n_{1}, n_{2})}(x^{\mu}) \nonumber \\ 
&&\varphi_{3}^{(-n_{1}, -n_{2})}(x^{\mu}) = - \varphi_{3}^{(n_{1}, n_{2})}(x^{\mu}).  
\label{4.4} 
\eea 

Instead of evaluating 2-point functions of $H_{1, 2}$ by directly calculating relevant Feynman diagrams, let us calculate 
the radiatively induced effective potential of $H_{1, 2}$ by the use of the background field method under the following background 
for the Higgs fields: 
\be
\label{4.5}
A^{(0, 0)}_{5, 6} =  \frac{1}{\sqrt{2}} 
\begin{pmatrix} 
0 & H_{1, 2} \\ 
H_{1, 2}^{\dagger} & 0 
\end{pmatrix}. 
\ee 
Under this background, gauge covariant derivatives along the extra space, 
when they act on the KK mode $(n_{1}, n_{2})$ of {\color{red} (\ref{4.3})}, are equivalent to the multiplications of the following matrices: 
\be 
\label{4.6} 
D_{5, 6} = i
\begin{pmatrix} 
\frac{n_{1,2}}{R}I_{2} &  \frac{g}{\sqrt{2}} H_{1,2} \\ 
\frac{g}{\sqrt{2}} H_{1,2}^{\dagger} & \frac{n_{1,2}}{R} 
\end{pmatrix}, 
\ee 
where $I_{2}$ is the $2 \times 2$ unit matrix. This leads to the ``mass-squared" operator for the KK mode,  
\be 
\label{4.7} 
{\cal M}_{n_{1}, n_{2}}^{2} = - (D_{5}^{2} + D_{6}^{2}) = \frac{n_{1}^{2}+n_{2}^{2}}{R^{2}} I_{3} 
+ 
\begin{pmatrix} 
 \frac{g^{2}}{2} (H_{1}H_{1}^{\dagger} + H_{2}H_{2}^{\dagger}) & \sqrt{2}g \frac{n_{1}H_{1} + n_{2}H_{2}}{R} \\ 
\sqrt{2}g \frac{n_{1}H_{1}^{\dagger} + n_{2}H_{2}^{\dagger}}{R} & \frac{g^{2}}{2}(H_{1}^{\dagger}H_{1} + H_{2}^{\dagger}H_{2})  
\end{pmatrix}. 
\ee 
We do not introduce a bulk mass for the scalar field, since the quadratic terms of the Higgs fields do not suffer from infrared divergences. 

The effective potential due to the bubble diagrams of the scalar matter fields is given as 
\be 
\label{4.8}
V_{eff}^{(s)} = \frac{1}{2} \int \frac{d^{4}p_{E}}{(2\pi)^{4}} \sum_{n_{1}=-\infty}^{\infty} \sum_{n_{2}=-\infty}^{\infty} 
{\rm Tr} \ \log (p_{E}^{2}I_{3} + {\cal M}_{n_{1}, n_{2}}^{2}), 
\ee 
where $I_{3}$ is the 3$\times$3 unit matrix and the Tr is taken over the 3$\times$3 matrix. 
$p_{E}$ is an Euclidean 4-momentum. The factor $\frac{1}{2}$ is to take care of the degeneracy (\ref{4.4}) 
[\,This prescription is also applicable to the KK zero-mode with $(n_{1},n_{2}) = (0, 0)$, since in (\ref{4.7}), 
the zero-mode contribution exists for the third component of the triplet, although actually the mode function for the third component, 
being an odd function, disappears for the KK zero mode\,].
If we ignore the charged scalars of $H_{1, 2}$, the three eigenvalues of the matrix ${\cal M}_{n_{1}, n_{2}}^{2}$ are 
\be 
\label{4.9}
\frac{n_{1}^{2}+n_{2}^{2}}{R^{2}}, \ \ \frac{n_{1}^{2}+n_{2}^{2}}{R^{2}} +\frac{g^{2}}{2}(|\phi_{1}^{0}|^{2} 
+ |\phi_{2}^{0}|^{2}) \pm \sqrt{2}g \frac{|n_{1}\phi_{1}^{0} + n_{2}\phi_{2}^{0}|}{R}. 
\ee 
The field-dependent eigenvalues cannot take a simple form, except in specific cases such as $\theta = 0$, where 
\be 
\label{4.10} 
\frac{n_{1}^{2}+n_{2}^{2}}{R^{2}} +\frac{g^{2}}{2}(|\phi_{1}^{0}|^{2} + |\phi_{2}^{0}|^{2}) \pm \sqrt{2}g \frac{|n_{1}\phi_{1}^{0} + n_{2}\phi_{2}^{0}|}{R} 
= \Bigl(\frac{n_{1}}{R} \pm \frac{g}{\sqrt{2}} |\phi_{1}^{0}| \Bigr)^{2} + \Bigl(\frac{n_{2}}{R} \pm \frac{g}{\sqrt{2}} |\phi_{2}^{0}| \Bigr)^{2}. 
\ee
The lesson here is that the evaluation of the whole effective potential by the use of Poisson resummation is difficult. 

Nevertheless, once we obtain the general formula (\ref{4.8}) for the effective potential, 
we easily get the quadratic terms of $H_{1,2}$ by the use of its Taylor expansion with respect to the fields $H_{1,2}$: 
\be 
\label{4.11}
V_{2}^{(s)} = \frac{1}{2} \int \frac{d^{4}p_{E}}{(2\pi)^{4}} \sum_{n_{1}, n_{2} =-\infty}^{\infty} 
\left[ \frac{g^{2}(H_{1}^{\dagger}H_{1} + H_{2}^{\dagger}H_{2})}{p_{E}^{2} + \frac{n_{1}^{2}+n_{2}^{2}}{R^{2}}} - 2g^{2} 
\frac{\frac{(n_{1}H_{1}^{\dagger} + n_{2}H_{2}^{\dagger})(n_{1}H_{1} + n_{2}H_{2})}{R^{2}}}{\bigl(p_{E}^{2} + \frac{n_{1}^{2}+n_{2}^{2}}{R^{2}}\bigr)^{2}}
\right]. 
\ee 
In the second term of the r.h.s. of (\ref{4.11}), the coefficient of the operator ${\rm Re}(H_{1}^{\dagger}H_{2})$ vanishes, 
just because it is proportional to $\sum_{n_{1}, n_{2} =-\infty}^{\infty} \frac{n_{1}n_{2}}{\bigl(p_{E}^{2} + \frac{n_{1}^{2}+n_{2}^{2}}{R^{2}}\bigr)^{2}} = 0$. 
Thus, by replacing $n_{1}^{2}H_{1}^{\dagger}H_{1}$ by $\frac{1}{2}(n_{1}^{2}+n_{2}^{2})H_{1}^{\dagger}H_{1}$ etc., 
invoking the symmetry between two extra spaces, (\ref{4.11}) is shown to yield only the operator $H_{1}^{\dagger}H_{1} + H_{2}^{\dagger}H_{2}$: 
\be 
\label{4.12} 
V_{2}^{(s)} = \frac{g^{2}}{2} \int \frac{d^{4}p_{E}}{(2\pi)^{4}} \sum_{n_{1}, n_{2} =-\infty}^{\infty} 
\left[ \frac{1}{p_{E}^{2} + \frac{n_{1}^{2}+n_{2}^{2}}{R^{2}}} - \frac{\frac{n_{1}^{2}+n_{2}^{2}}{R^{2}}}{\bigl(p_{E}^{2} + \frac{n_{1}^{2}+n_{2}^{2}}{R^{2}}\bigr)^{2}} \right] 
(H_{1}^{\dagger}H_{1} + H_{2}^{\dagger}H_{2}).  
\ee    
Thus in this model, there is no quadratic term other than the terms with coefficients $a$ and $b$ in (\ref{3.8'}), 
as we expected, and the contributions to these coefficients due to the matter scalars are given as 
\bea 
&& a^{(s)} = \frac{g^{2}}{2} \int \frac{d^{4}p_{E}}{(2\pi)^{4}} \sum_{n_{1}, n_{2} =-\infty}^{\infty} 
\left[ \frac{1}{p_{E}^{2} + \frac{n_{1}^{2}+n_{2}^{2}}{R^{2}}} - \frac{\frac{n_{1}^{2}+n_{2}^{2}}{R^{2}}}{\bigl(p_{E}^{2} + \frac{n_{1}^{2}+n_{2}^{2}}{R^{2}}\bigr)^{2}} \right], \nonumber \\ 
&& b^{(s)} = 0.
\label{4.13} 
\eea   
Utilizing the formula 
\be 
\label{4.14}
\frac{1}{\alpha} = \int_{0}^{\infty} \ e^{-\alpha t} dt, \ \ \frac{1}{\alpha^{2}} = \int_{0}^{\infty} \ te^{-\alpha t} dt,  
\ee
we get 
\be 
\label{4.15} 
a^{(s)} =  \frac{g^{2}}{2} \int_{0}^{\infty} \ dt \int \frac{d^{4}p_{E}}{(2\pi)^{4}} \sum_{n_{1}, n_{2} =-\infty}^{\infty}  
\Bigl(1 - t \frac{n_{1}^{2}+n_{2}^{2}}{R^{2}}\Bigr) e^{- \bigl(p_{E}^{2} + \frac{n_{1}^{2}+n_{2}^{2}}{R^{2}}\bigr) t}.
\ee 
Then, a manipulation by the use of the Poisson resummations [\,$k_{1}$ and $k_{2}$ are winding numbers\,],  
\bea  
\sum_{n_{1}, n_{2}} e^{-t \frac{n_{1}^{2}+n_{2}^{2}}{R^{2}}} &=& \pi R^{2} \frac{1}{t} \sum_{k_{1}, k_{2}} e^{-\frac{(\pi R)^{2}(k_{1}^{2} + k_{2}^{2})}{t}}, \nonumber \\ 
\sum_{n_{1}, n_{2}} \frac{n_{1,2}^{2}}{R^{2}} e^{-t \frac{n_{1}^{2}+n_{2}^{2}}{R^{2}}} &=& \pi R^{2} \sum_{k_{1}, k_{2}} 
\Bigl(\frac{1}{2t^{2}} - \frac{(\pi R)^{2}}{t^{3}}k_{1,2}^{2}\Bigr) e^{-\frac{(\pi R)^{2}(k_{1}^{2} + k_{2}^{2})}{t}}, 
\label{4.16}  
\eea
leads to 
\bea 
a^{(s)} &=&  \frac{g^{2}}{2} \pi R^{2} \int_{0}^{\infty} \ dt \int \frac{d^{4}p_{E}}{(2\pi)^{4}} 
\sum_{k_{1}, k_{2} =-\infty}^{\infty} \left[\frac{1}{t} - \Bigl(\frac{1}{t} - \frac{(\pi R)^{2}}{t^{2}} (k_{1}^{2} + k_{2}^{2}) \Bigr) \right] 
e^{- t p_{E}^{2}} e^{-\frac{(\pi R)^{2}(k_{1}^{2} + k_{2}^{2})}{t}} \nonumber \\ 
&=& \frac{\pi^{3}g^{2}}{2} R^{4} \int_{0}^{\infty} \ \frac{dt}{t^{2}} \int \frac{d^{4}p_{E}}{(2\pi)^{4}} 
\sum_{k_{1}, k_{2} =-\infty}^{\infty} (k_{1}^{2} + k_{2}^{2}) e^{- t p_{E}^{2}} e^{-\frac{(\pi R)^{2}(k_{1}^{2} + k_{2}^{2})}{t}}. 
\label{4.17}  
\eea 

First, we comment on the zero-winding sector, $(k_{1}, k_{2}) = (0, 0)$, 
which corresponds to the contribution to the local mass-squared operator of the Higgs fields and should be forbidden by local gauge symmetry. 
In fact, (\ref{4.17}), having the prefactor $k_{1}^{2} + k_{2}^{2}$, clearly implies that the contribution of the zero-winding sector vanishes. 
Thus, as we expected, the zero-winding sector disappears, and in the remaining contribution the integrals over $t$ and 
4-momentum $p_{E}$ are convergent. Then, using the formula 
\be 
\label{4.18} 
\int \frac{d^{4}p_{E}}{(2\pi)^{4}} \ e^{- t p_{E}^{2}} = \frac{1}{16\pi^{2}}\frac{1}{t^{2}},  
\ee 
we get  
\be 
\label{4.19}
a^{(s)} =  \frac{\pi g^{2}}{32} R^{4} \int_{0}^{\infty} \ \frac{dt}{t^{4}} \sum_{(k_{1}, k_{2}) \neq (0, 0)} (k_{1}^{2} + k_{2}^{2}) e^{-\frac{(\pi R)^{2}(k_{1}^{2} + k_{2}^{2})}{t}}. 
\ee 
By changing the variable, $\frac{(\pi R)^{2}(k_{1}^{2} + k_{2}^{2})}{t} \to l$, and performing the integral over $l$, we finally get 
\bea  
&&a^{(s)} =  \frac{g^{2}}{16\pi^{5}} \frac{1}{R^{2}} \sum_{(k_{1}, k_{2}) \neq (0, 0)} \frac{1}{(k_{1}^{2} + k_{2}^{2})^{2}} = 5.3 \times 10^{-4}\frac{1}{R^{2}} , \nonumber \\ 
&&b^{(s)} = 0. 
\label{4.20} 
\eea 

\subsection{A model with matter fermion} 

In the model with scalar matter fields, although the necessary condition (\ref{3.9'}) is satisfied, 
unfortunately another condition (\ref{3.15}) is not satisfied, as we see in (\ref{4.20}).   

Thus, we now discuss a model with SU(3) triplet fermions as the matter fields: 
\be 
\label{5.1} 
\Psi = 
\begin{pmatrix} 
\psi_{1} \\ 
\psi_{2} \\ 
\psi_{3} 
\end{pmatrix}. 
\ee 
The quantum correction due to the matter fermions is known to yield the nonvanishing coefficient $b$ 
through the commutator of gauge co variant derivatives, $[D_{5}, D_{6}]$, as we will see below. 

The 6D gamma matrices are given in the space of the direct product of the 4D spinor space 
and [\,SU(2)\,] internal space as 
\be  
\Gamma^{\mu} = \gamma^{\mu} \otimes I_2, \ \ \Gamma^5 = \gamma^5 \otimes i \sigma_1 , \ \ 
\Gamma^6 = \gamma^5 \otimes i \sigma_2 \ \ \ (\mu = 0, 1, 2, 3).  
\label{5.2}
\ee 
Then, the 6D chiral operator is given as 
\be 
\label{5.3} 
\Gamma_{7} = \Gamma^{0} \Gamma^{1} \Gamma^{2} \Gamma^{3} \Gamma^{5} \Gamma^{6} 
= - \gamma^{5} \otimes \sigma_3 
\ee 
Our matter fermion is assumed to be 6D Weyl fermion: 
\be 
\label{5.4}
\Gamma_{7} \Psi = - \Psi. 
\ee 
Let us note that, even if we adopt the 6D Weyl fermion with the eigenvalue $-1$ of $\Gamma_{7}$, 
there are two cases, 
4D right-handed fermion with +1 eigenvalue of $\sigma_{3}$
and
4D left-handed fermion with $-1$ eigenvalue of $\sigma_{3}$. 

The $Z_2$-parity assignment for $\Psi$ is 
\be 
\label{5.5} 
\Psi (-x^{5}, -x^{6}) = P(-i\Gamma_{4} \Gamma_{5}) \Psi (x^{5}, x^{6}) = - P  I_{4} \otimes \sigma_3 \Psi (x^{5}, x^{6}). 
\ee 
Thus, its KK mode expansion is as follows: 
\be 
\label{5.6}
\Psi (x^{\mu}, x^{5}, x^{6}) = \sum_{n_{1}=-\infty}^{\infty} \sum_{n_{2}=-\infty}^{\infty} \frac{1}{2 \pi R} 
\begin{pmatrix} 
\cos ( \frac{n_{1}x^{5} + n_{2}x^{6}}{R} ) \psi_{1L}^{(n_{1}, n_{2})}(x^{\mu}) + i\sin ( \frac{n_{1}x^{5} + n_{2}x^{6}}{R} ) \psi_{1R}^{(n_{1}, n_{2})}(x^{\mu}) \\ 
\cos ( \frac{n_{1}x^{5} + n_{2}x^{6}}{R} ) \psi_{2L}^{(n_{1}, n_{2})}(x^{\mu}) + i\sin ( \frac{n_{1}x^{5} + n_{2}x^{6}}{R} ) \psi_{2R}^{(n_{1}, n_{2})}(x^{\mu}) \\ 
i \sin ( \frac{n_{1}x^{5} + n_{2}x^{6}}{R} ) \psi_{3L}^{(n_{1}, n_{2})}(x^{\mu}) + \cos ( \frac{n_{1}x^{5} + n_{2}x^{6}}{R} ) \psi_{3R}^{(n_{1}, n_{2})}(x^{\mu}) 
\end{pmatrix},  
\ee 
where there are degeneracies among the 4D fields because of the $Z_{2}$ orbifolding:   
\bea 
&&\psi_{1L, 2L}^{(-n_{1}, -n_{2})}(x^{\mu}) = \psi_{1L, 2L}^{(n_{1}, n_{2})}(x^{\mu}), \ \ \psi_{3R}^{(-n_{1}, -n_{2})}(x^{\mu}) = \psi_{3R}^{(n_{1}, n_{2})}(x^{\mu}) \nonumber \\ 
&&\psi_{1R, 2R}^{(-n_{1}, -n_{2})}(x^{\mu}) = - \psi_{1R, 2R}^{(n_{1}, n_{2})}(x^{\mu}), \ \ \psi_{3L}^{(-n_{1}, -n_{2})}(x^{\mu}) = - \psi_{3L}^{(n_{1}, n_{2})}(x^{\mu}).  
\label{5.7} 
\eea  

Similarly to the case of the model with a matter scalar, the mass-squared matrix, i.e., the squared Dirac operator, is calculated to be 
\bea 
&&\tilde{{\cal M}}_{n_{1}, n_{2}}^{2} =  (D_{5}\Gamma_{5} + D_{6}\Gamma_{6})^{2} = \sum_{a, b = 5, 6} \Gamma_{a}\Gamma_{b}D_{a}D_{b} 
\nonumber \\ 
&&= \sum_{a, b = 5, 6} \Bigl\{ \frac{1}{2}\{\Gamma_{a}, \Gamma_{b}\} D_{a}D_{b} + \frac{1}{4}[\Gamma_{a}, \Gamma_{b}] [D_{a}, D_{b}]\Bigr\}  \nonumber \\ 
&&= {\cal M}_{n_{1}, n_{2}}^{2} - g^{2} \Gamma_{5} \Gamma_{6} [A_{5}, A_{6}] \nonumber \\ 
&&= {\cal M}_{n_{1}, n_{2}}^{2} + \frac{i}{2}g^{2} (I_{4} \otimes \sigma_{3}) \cdot 
\begin{pmatrix} 
H_{1}H_{2}^{\dagger} - H_{2}H_{1}^{\dagger} & 0 \\ 
0 & H_{1}^{\dagger}H_{2} - H_{2}^{\dagger}H_{1} 
\end{pmatrix}.  
\label{5.8}  
\eea 
We naively expect that the additional operator with the prefactor $I_{4} \otimes \sigma_{3}$, 
the contribution of the  commutator $[D_{5}, D_{6}]$, i.e., of the ``tadpole" $F_{5, 6}$, vanishes under the Tr in the evaluation 
of the effective potential. 
In fact, for nonzero KK modes, each component of $\Psi$ has both 4D chiralities, 
i.e., both $\pm 1$ eigenvalues of $\sigma_{3}$ and the sum of the eigenvalues of the additional operator just vanishes. 
For the KK zero-mode sector, however, $\psi_{1, 2}$ and $\psi_{3}$ are L and R 4D Weyl spinors.
This means that $\psi_{1, 2}$ and $\psi_{3}$ have $-1$ and $+1$ eigenvalues of $\sigma_{3}$, respectively, and the sum of 
the eigenvalues is nonvanishing\,:
\be 
\label{5.8'}
{\rm Tr}\Bigl[ 
\begin{pmatrix} 
- I_{2} & 0 \\ 
0 & 1
\end{pmatrix}
\cdot 
\begin{pmatrix} 
H_{1}H_{2}^{\dagger} - H_{2}H_{1}^{\dagger} & 0 \\ 
0 & H_{1}^{\dagger}H_{2} - H_{2}^{\dagger}H_{1} 
\end{pmatrix} \Bigr] 
= 2(H_{1}^{\dagger}H_{2} - H_{2}^{\dagger}H_{1}). 
\ee
Thus, we will take only the KK zero-mode into account when we evaluate the tadpole term.  

The effective potential due to the bubble diagram of the matter fermion is given similarly to (\ref{4.8}) as 
\be 
\label{5.9}
V_{eff}^{(f)} = - \frac{1}{2}\times 2 \int \frac{d^{4}p_{E}}{(2\pi)^{4}} \sum_{n_{1}=-\infty}^{\infty} \sum_{n_{2}=-\infty}^{\infty} 
{\rm Tr} \ \log (p_{E}^{2}I_{3} + \tilde{{\cal M}}_{n_{1}, n_{2}}^{2}).  
\ee 
Again, we easily get the quadratic terms of $H_{1, 2}$ by performing the Taylor expansion with respect to $H_{1, 2}$. 
The contribution of ${\cal M}_{n_{1}, n_{2}}^{2}$ in $\tilde{{\cal M}}_{n_{1}, n_{2}}^{2}$ is exactly the same as 
that in the case of the model with a scalar matter, 
except for the difference in the overall factor, and the additional tadpole contribution in $\tilde{{\cal M}}_{n_{1}, n_{2}}^{2}$ 
readily leads to the coefficient $b$. 
Namely, the fermionic contributions to the coefficients $a$ and $b$ in (\ref{3.8'}) are given as 
\bea 
&& a^{(f)} =  - \frac{g^{2}}{8\pi^{5}} \frac{1}{R^{2}} \sum_{(k_{1}, k_{2}) \neq (0, 0)} \frac{1}{(k_{1}^{2} + k_{2}^{2})^{2}} = - 1.1\times 10^{-3}\frac{1}{R^{2}},  \nonumber \\
&& b^{(f)} = 2g^{2} \int \frac{d^{4}p_{E}}{(2\pi)^{4}} \frac{1}{p_{E}^{2}}.  
\label{5.10}
\eea 
We realize that $b^{(f)}$ is quadratically UV-divergent, as we anticipated, 
and the condition (\ref{3.15}) is easily satisfied, although we need a renormalization procedure. 
We also note that the sign of $a^{(f)}$ is opposite to that of $a^{(s)}$, coming from the difference in the statistics. 
Thus, supposing we introduce $n_{s}$ matter scalars and $n_{f}$ matter fermions, the condition (\ref{3.9'}) 
requires [\,see (\ref{4.20}) and (\ref{5.10})\,] 
\be 
\label{5.11}
n_{s} - 2n_{f} > 0.
\ee 
We, however, should note that to make the analysis more realistic, 
we anyway need to take the quantum corrections due to the gauge fields and Higgs fields into account, 
in addition to the contribution due to the matter fields. 
In fact, the contributions to the coefficient $a$ by such bosonic states are expected to be positive, 
and the condition (\ref{3.9'}) may be satisfied without introducing any matter scalar fields. 

Now, one comment is in order. 
Similar discussions concerning the effective Higgs potential and mass eigenvalues of physical scalars 
in the 6D U(3) GHU model with two Higgs doublets already exist in the literature \cite{ABQ, HNT}. 
In these works, however, the effective potential was evaluated only for the flat direction, 
$H_{1} \propto H_{2}$, and therefore the obtained vacuum states are different from what we obtained in this work. 
However, the mass eigenvalues of physical scalars were discussed by considering the fluctuations of the scalar fields 
around the origin or the vacuum state along the flat direction.

\section{SU(4) GHU Model with Two Higgs Doublets} 

Although the 6D SU(3) GHU model with two Higgs doublets has an attractive feature that the predicted Higgs mass 
satisfies $M_{H} \leq 2M_{W}$ at the leading order of the perturbative expansion and even coincides with the 
observed value for a suitable choice of the parameter $a$, the predicted weak mixing angle is unrealistic: $\sin^{2} \theta_{W} = 3/4$. 

In this section, we very briefly discuss a model that can possibly account for both the observed Higgs mass 
and a realistic weak mixing angle, $\sin^{2} \theta_{W} = 1/4$, 
by the use of a familiar unitary gauge group. 
The model is the 6D SU(4) GHU model on the $T^{2}/Z_2$ orbifold as the extra space. 
Because of the $Z_2$ orbifolding, the model now involves two Higgs doublets behaving as triplets of the SU(3) subgroup 
[\,Refer to the discussion in Section 4 for the SU(4) model with one Higgs doublet\,].
In fact, by a suitable assignment of $Z_2$-parities, we realize that the KK zero-modes of $A_{5, 6}$ behave as $3 + \bar{3}$, 
not $8$ of SU(3), as we will see below, thus leading to the successful prediction of the weak mixing angle. 

What we need to realize for the KK zero-modes of the gauge-Higgs sector are the following forms: 
\bea 
&&A_{\mu} =  
\begin{pmatrix} 
\frac{1}{2}\gamma_{\mu} - \frac{\sqrt{3}}{6}Z_{\mu} & \frac{1}{\sqrt{2}} W^{+}_{\mu} & 0 & 0 \\ 
\frac{1}{\sqrt{2}} W^{-}_{\mu} & \frac{\sqrt{3}}{3}Z_{\mu} & 0 & 0 \\ 
0 & 0 & - \frac{1}{2}\gamma_{\mu} - \frac{\sqrt{3}}{6}Z_{\mu} & 0 \\ 
0 & 0 & 0 & 0 
\end{pmatrix} + X_{\mu}\frac{\sqrt{6}}{12} \ {\rm diag} \ (1, 1, 1, -3),   
\label{6.2a} \\ 
&&A_{5, 6} = \frac{1}{\sqrt{2}} 
\begin{pmatrix} 
0 & 0 & 0 & \phi_{1, 2}^{+} \\ 
0 & 0 & 0 & \phi_{1, 2}^{0} \\
0 & 0 & 0 & 0 \\ 
\phi_{1, 2}^{-} & \phi_{1, 2}^{0\ast} & 0 & 0 
\end{pmatrix},  
\label{6.2b} 
\eea 
where $X_{\mu}$ is the gauge boson of the extra U(1)$_X$. 

The idea to realize the KK zero-modes shown above is to break SU(4) in two steps, 
SU(4) $\to$ SU(3)$\times$ U(1)$_X \to$ SU(2)$_{L} \times$ U(1)$_{Y} \times$ U(1)$_X$, 
by assigning different $Z_2$-parities $(+, +, +, -), \ (+, +, -, -)$,
where
$+$ and $-$ stands for $+1$ and $-1$, respectively, for the reflection at two different fixed points, 
$(x^{5}, \ x^{6}) = (0, \ 0), \ (\pi R, \pi R)$, concerning the fundamental repr. of SU(4). 
Namely, we assign $Z_2$ parities under 
the rotation around
two fixed points for the fundamental repr. as 
\be 
\label{6.3}
\begin{pmatrix} 
(+, +) \\ 
(+, +) \\ 
(+, -) \\ 
(-, -)  
\end{pmatrix}. 
\ee 
Accordingly, the  $Z_2$ parity assignments for the 4D gauge bosons and 4D scalars, i.e., the Higgs fields, are given as 
\bea 
&&A_{\mu} =  
\begin{pmatrix} 
(+, +) & (+, +) & (+, -) & (-, -) \\ 
(+, +) & (+, +) & (+, -) & (-, -) \\ 
(+, -) & (+, -) & (+, +) & (-, +) \\ 
(-, -) & (-, -) & (-, +) & (+, +) 
\end{pmatrix},   
\label{6.4a} \\ 
&&A_{5, 6} =  
\begin{pmatrix} 
(-, -) & (-, -) & (-, +) & (+, +) \\ 
(-, -) & (-, -) & (-, +) & (+, +) \\ 
(-, +) & (-, +) & (-, -) & (+, -) \\ 
(+, +) & (+, +) & (+, -) & (-, -) 
\end{pmatrix}.  
\label{6.4b} 
\eea   

The calculation of the effective potential $V (H_{1}, H_{2})$ of the two Higgs doublets is exactly the same as in the case 
of the SU(3) model, 
discussed in the previous section; therefore, the relation $M_{H} < 2M_{W}$ can be realized 
by introducing matter fermions belonging to the fundamental repr. of SU(4), 
\be 
\label{6.5} 
\Psi = 
\begin{pmatrix} 
\psi_{1} \\ 
\psi_{2} \\ 
\psi_{3} \\ 
\psi_{4}
\end{pmatrix},  
\ee 
whose components are given $Z_2$-parities, in the same way as (\ref{5.5}) with 
\be 
\label{6.6}
P = {\rm diag} \ ((1, 1), \ (1, 1), \ (1, -1), \ (-1, -1)),  
\ee 
in accordance with (\ref{6.3}). 
Now a pair $(\psi_{2}, \psi_{4})$ couples with $\phi_{1, 2}^{0}$ and corresponds to the pair $(\psi_{2}, \psi_{3})$ in the SU(3) model.

\section{Summary} 

We discussed the scenario of gauge-Higgs unification (GHU), 
where the Higgs field is identified with the extra space component of the higher dimensional gauge field, as an 
interesting candidate of BSM physics. 
GHU models with multi-dimensional extra space predict at the classical level the Higgs self-coupling $\lambda \sim g^{2}$ 
and a light Higgs with the mass of ${\cal O}(M_{W})$, similarly to the case of MSSM. 

It has been known that the 6D SU(3) GHU model with one Higgs doublet in its low energy effective theory 
predicts an interesting relation $M_{H} = 2M_{W}$ at the leading order of the perturbative expansion \cite{SSSW, LMM}. 
In our previous paper \cite{LMM}, 
we demonstrated that the ratio of the Higgs mass to the weak scale is calculable as a UV-finite value even under the quantum correction 
and it is possible to recover the observed Higgs mass, $M_{H} = 125$ GeV. 
We, however, noticed that, to realize the difference in $2M_{W}$ and the observed mass,
a rather large quantum correction is needed.  

There is another interesting prediction of the GHU scenario, namely, the prediction of the weak mixing angle,
i.e., the mass ratio of weak gauge bosons. 
Unfortunately, in the minimal SU(3) GHU model with a simple gauge group \cite{SSSW, KLY}, the predicted weak mixing angle is unrealistic: $\sin^{2}\theta_{W} = 3/4$.    

We thus addressed a question whether there ever exist GHU models that provide more realistic predictions of the Higgs mass 
and the weak mixing angle at the leading order of the perturbative expansion. 
Namely, we investigated in the scheme of 6D GHU whether the predictions of the Higgs mass 
and the weak mixing angle become closer to or coincide with the observed values by suitable choices of 
the gauge group and the orbifolding of the extra space.  

We first discussed the weak mixing angle. By the use of the useful formula (\ref{0.3}), 
we studied which repr. of the minimal group SU(3), which is a subgroup of the adopted gauge group in general, 
the Higgs doublet should belong to in order to realize the realistic prediction, $\sin^{2}\theta_{W} = 1/4$. 
We showed that, among the repr.s up to the 2nd rank tensor, 
triplet
and sextet repr.s of SU(3) lead to the realistic prediction. 
The decomposition of the adjoint repr. of $G_2$ under the subgroup SU(3) contains the triplet repr. 
and this is why the gauge group can predict $\sin^{2}\theta_{W} = 1/4$ \cite{1979Manton, CGM}.

Next, we investigated the 6D Sp(6) GHU model with one Higgs doublet, as a prototype model whose adjoint repr. contains 
the sextet repr. of SU(3), 
the new possibility to get the realistic weak mixing angle. 
We have found that, although the weak mixing angle and the mass ratio of the weak gauge bosons are confirmed to be realistic as we expected, 
the predicted Higgs mass  satisfies $M_{H} = 2M_{W}$, just as in the case of the 6D SU(3) GHU model with one Higgs doublet. 
We also briefly investigated the 6D SU(4) model with one Higgs doublet, as another possibility containing the triplet repr. of 
SU(3) with the familiar unitary gauge group. 
We again found that the predicted Higgs mass is $2M_{W}$, while keeping the successful weak mixing angle. 

In Sections 5 and 6, we discussed 6D GHU models with two Higgs doublets taking the choice of $Z_2$ orbifolding for the extra space, 
hoping that the prediction of the Higgs mass becomes more realistic than those in the models with one Higgs doublet. 
As the minimal model for such a purpose, in Section 5, we first investigated the 6D SU(3) model with two Higgs doublets in some detail. 
We first gave a general argument on the form of the effective potential for two Higgs doublets $H_{1, 2}$ 
relying on the higher dimensional gauge symmetry and the symmetry of the torus as the extra space. 
After the minimization of the potential, we calculated the mass eigenvalues for the five physically remaining scalar particles. 
Among other things, we have found that the prediction of the Higgs mass at the leading order of the perturbation is 
\be 
\label{7.1} 
M_{H} \leq 2M_{W};
\ee 
therefore, it is possible to realize the observed 125 GeV for a suitable choice of the parameter $a$ in the potential (\ref{3.8'}), 
which is calculable as a function of $R$, the size of the extra space, and the gauge coupling $g$. 
This is in clear contrast to the case of MSSM, 
where the Higgs mass satisfies $M_{H} \leq M_{Z} \cos 2\beta \leq M_{Z}$ at the classical level and has no chance
of being in agreement with the observed value.  

Both the parameters $a$ and $b$ in the potential (\ref{3.8'}) are radiatively induced 
and we performed concrete calculations of the quantum corrections using the background field method. 
We realized that the parameter $b$, which corresponds to the contribution of the ``tadpole" term localized at the fixed points 
and plays an important role in making the model realistic, is induced only in the theory with fermions through the commutator of covariant derivatives $[D_{5}, \ D_{6}]$. 

In Section 6, we very briefly investigated the 6D SU(4) model with two Higgs doublets, for the purpose of improving
the prediction of the weak mixing angle, 
while keeping the interesting feature $M_{H} \leq 2M_{W}$ obtained from the general argument of the Higgs potential in Section 5.

An important central issue in the GHU scenario is how to generate the necessary hierarchy between the weak scale $M_{W}$ and the compactification scale $M_{c} = \frac{1}{R}$  ($R$: the size of the extra dimension). 
In the simplest 5D GHU models (formulated on a flat space-time), the radiatively induced Higgs potential, being described by the Wilson-loop phase, 
is completely UV-finite and also periodic in the Higgs field with a period $\sim \frac{1}{gR}$. 
Thus, by writing the Higgs VEV as $v = \frac{\alpha}{gR}$ with a 
parameter $\alpha$,  $\alpha$ is usually of ${\cal O}(1)$ 
and the weak scale $M_{W} \sim gv = \frac{\alpha}{R}$ is comparable to $\frac{1}{R}$. 
This means that $M_{c} \sim M_{W}$, unless $\alpha$ becomes small for some reason 
(e.g., by the introduction of many exotic matter fields belonging to the  adjoint repr. of the gauge group \cite{KLY}), 
and will lead to an immediate contradiction with the recent data from LHC experiments, which have not seen any evidence of BSM physics. 

In the GHU models with multiple extra dimensions discussed in this paper, however, the situation is different. 
Namely, although the ratio $\frac{M_{H}}{M_{W}}$ is predictable to be finite,
we realize that $M_{W}$ itself is not correlated with $M_{c}$. 
The essential difference from the case of 5D GHU is that the radiatively induced VEV $v$ or the weak scale $M_{W}$
is UV-divergent due to the divergent coefficient $b$, as seen in (\ref{3.17}) and (\ref{3.18}). 
Hence, the VEV and therefore $M_{W}$ need to be renormalized and are not calculable. 
Now, $M_{c}$ is fixed so that it recovers the observed Higgs mass: as is seen in (\ref{3.53}), (\ref{4.20}), and (\ref{5.10}) 
the observed Higgs mass is realized for $M_{c} = \frac{1}{R}$ of a few TeV or so, not $M_{W}$.
Note that the UV divergence in the quadratic term in the Higgs potential has its origin in the 
fact that our vacuum state is not along the flat direction: $[\langle H_{1} \rangle, \langle H_{2} \rangle] \neq 0$.

This work is the first step toward a truly realistic GHU model with successful mass ratios of the Higgs boson and weak gauge bosons, 
and there remain issues to be settled. Concerning the prediction of the weak mixing angle, 
since Sp(6) and SU(4) are groups of rank 3, the unnecessary additional U(1) gauge boson needs to be removed. 
This may be realized by invoking the anomaly of the associated gauge symmetry or by putting an additional mass term for
the gauge boson at the orbifold fixed points. 

We also note that $q = 1$ and $0$, necessary to realize the realistic prediction $\sin^{2}\theta_{W} = 1/4$, 
implies that all the components of the SU(3) triplet have integer charges as seen from (\ref{0.2}). 
This means that all the components of the repr.s of the theory have integer charges, and quarks cannot be 
assigned to any repr. of the bulk gauge symmetry. 
Then, we encounter the problem on how the quark fields are incorporated to the theory. 
One possibility may be to put the quark fields on the fixed points of the orbifold, as was proposed in \cite{CGM}. 

Concerning the prediction of the Higgs mass, the values of the coefficients $a$ and $b$ in the potential (\ref{3.8'}), 
which play crucial roles in the prediction of the Higgs mass and to realize the spontaneous symmetry breaking, 
are sensitive to the content of the fields contributing to the quantum corrections of $a$ and $b$. 
Thus, we need a full calculation of the quantum correction including the contributions of the gauge-Higgs sector in 
addition to that of the matter fields, 
before we get a conclusive prediction of the Higgs mass as the function of the compactification scale. 
On the other hand, we expect that the desirable feature (\ref{7.1}) of the GHU model itself, derived from the argument based on the general form of the effective potential, will not change. 

Finally, in the attractive case of $M_{H} < 2M_{W}$, the identified Higgs field $\tilde{h}$ is not the field developing the VEV $v$, 
although it has Yukawa couplings with matter fermions through higher-dimensional gauge interaction. 
Thus, the Higgs boson does not behave like the one in the SM. Let us note that, also in MSSM, 
the lighter CP-even state does not coincide with the field developing the VEV. 
Thus, the situation mentioned above in our two doublet models may change by introducing the degrees of freedom, 
corresponding to the two angles $\alpha$ and $\beta$ (denoting the relative weights of two doublets $H_{U}$ and $H_{D}$
in the CP-even mass eigenstates and the VEV) in the MSSM. 
Such modification of the model will be possible if we adopt ``asymmetric torus" with $|\vec{l}_{1}| \neq |\vec{l}_{2}| $ 
and/or an arbitrary relative angle between two lattice vectors $\vec{l}_{1}$ and $\vec{l}_{2}$, as was discussed in 
the literature \cite{ABQ, HNT}.   

We leave these issues for a future publication.

\subsection*{Acknowledgments}
This work of C.S.L. was supported in part by the Grant-in-Aid for Scientific Research 
of the Ministry of Education, Science and Culture, No.s 15K05062 and 23104009.


\end{document}